\documentclass[12pt,preprint]{aastex}

\shorttitle{Galactic FUV Extinction}
\shortauthors{Sofia et al.}

\begin{document}

\title{{\it FUSE} Measurements of Far Ultraviolet Extinction. I. Galactic Sight Lines\footnote{Based on
observations with the NASA-CNES-CSA {\it Far Ultraviolet Spectroscopic Explorer}, which is operated by the Johns Hopkins University under NASA contract
NAS 32985.}}

\author{Ulysses J. Sofia\footnote{sofiauj@whitman.edu}}
\affil{Astronomy Department, Whitman College,
    Walla Walla, WA 99362}

\author{Michael J. Wolff\footnote{wolff@spacescience.org}}
\affil{Space Science Institute, 
Boulder, CO, 80303}

\author{Brian Rachford\footnote{rachford@casa.colorado.edu}}
\affil{Center for Astrophysics and Space Astronomy, Department of 
Astrophysical and Planetary Sciences, University of Colorado
Boulder, CO, 80309}

\author{Karl D. Gordon\footnote{kgordon@as.arizona.edu}}
\affil{Steward Observatory, University of Arizona 
Tucson, AZ 85721}

\author{Geoffrey C. Clayton\footnote{gclayton@fenway.phys.lsu.edu} \ 
and Stefan I. B. Cartledge\footnote{cartledg@taurus.phys.lsu.edu}}
\affil{Department of Physics and Astronomy, Louisiana State University, 
Baton Rouge, LA 70803}

\author{Peter G. Martin\footnote{pgmartin@cita.utoronto.ca}}
\affil{Canadian Institute for Theoretical Astrophysics, University of Toronto 
Toronto, Ontario, Canada M5S 3H8}

\author{Bruce T. Draine\footnote{draine@astro.princeton.edu}}
\affil{Princeton University Observatory, 
Princeton, NJ  08544}

\author{John S. Mathis\footnote{mathis@astro.wisc.edu}}
\affil{Department of Astronomy, University of Wisconsin 
Madison, WI 53706}

\author{Theodore P. Snow\footnote{tsnow@casa.colorado.edu}}
\affil{Center for Astrophysics and Space Astronomy, Department of 
Astrophysical and Planetary Sciences, University of Colorado
Boulder, CO, 80309}

\author{Douglas C. B. Whittet\footnote{whittd@rpi.edu}}
\affil{Department of Physics and Astronomy, Rensselaer Polytechnic Institute
Troy, NY 12180}

\begin{abstract}

We present extinction curves that include data down to far ultraviolet 
wavelengths (FUV; 1050 -- 1200 \AA) for nine Galactic sight lines. The 
FUV extinction was measured using data from the {\it Far Ultraviolet 
Spectroscopic Explorer}. The sight lines were chosen for their unusual 
extinction properties in the infrared through the ultraviolet; that they 
probe a wide range of dust 
environments is evidenced by the large spread in their measured ratios 
of total-to-selective extinction, $R_{V}$ = 2.43 -- 3.81. We find that 
extrapolation of the Fitzpatrick \& Massa relationship from the ultraviolet 
appears to be a good predictor of the FUV extinction behavior. 
We find that predictions of the FUV extinction based
upon the Cardelli, Clayton \& Mathis (CCM) dependence on $R_{V}$ give
mixed results. For the seven 
extinction curves well represented by CCM in the infrared through ultraviolet
(x $<$ 8~\micron$^{-1}$), the 
FUV extinction is well predicted in three sight lines, over-predicted 
in two sight lines, and under-predicted in 2 sight lines. A Maximum 
Entropy Method analysis using a simple three component grain model
shows that seven of the nine sight lines in 
the study require a larger fraction of grain materials to be in dust 
when FUV extinction is included in the models. Most of the added grain 
material is in the form of small (radii $\lesssim$ 200 \AA) grains.
\end{abstract}


\section{Introduction}

Little is known about Galactic extinction in the far ultraviolet region of 
the spectrum (FUV; 912 -- 1150 \AA\ or 8.7 - 11.0 $\mu$m$^{-1}$; in this
paper we include wavenumbers from 3.3 -- 8.7 $\mu$m$^{-1}$ in what we refer 
to as the UV since {\it IUE} covered this spectral region) other than 
that it is relatively strong, it increases with increasing wavenumber, and is 
variable among sight lines. The strength of FUV extinction requires that 
it be properly considered when recovering the intrinsic spectrum of an 
object. This is true for most observations taken in the FUV as well as for 
ultraviolet (UV; 1150 -- 3300 \AA\ or 3.3 -- 8.7 \micron$^{-1}$) or
optical observations of significantly red-shifted 
objects. The light 
extinguished at FUV wavelengths plays a major role in the energetics of 
star-forming galaxies since a large fraction of energy
in starlight is processed to 
long wavelengths through absorption and re-emission by dust. Molecular 
cloud physics and chemistry are also greatly affected by the amount of 
UV and FUV radiation that penetrates such regions; it is these 
wavelengths that will most affect the formation and destruction of molecules. 
Finally, the strength and variation of FUV extinction make it a valuable 
diagnostic for better characterizing the mass of material in and the
compositions of small grains ($\lesssim 200$ \AA).

Extinction curves in the ultraviolet (3.3 --  8.7 $\mu$m$^{-1}$) have 
been well approximated by two empirical relationships. For the first of 
these, Fitzpatrick \& Massa (1986; 1988; 1990, hereafter FM) found an 
expression for extinction that relies on six parameters, two that account 
for an underlying linear extinction ($c_{1}$ and $c_{2}$), three that 
describe the strength, location, and width of the 2175 \AA\ bump ($c_{3}$, 
$x_{0}$ and $\gamma$) and one that describes the curvature of the rise 
toward the FUV ($c_{4}$). These parameters are simply coefficients
in mathematical functions that were fit to UV data, and have no known 
physical basis. \citet{FIT99} argues that the location 
of the 2175 \AA\  bump is stable and that the underlying linear extinction 
can be varied with a single parameter, so that FM fits can be well determined 
with only four parameters. The consistent position of the 2175 \AA\ feature 
is confirmed by 
\citet{VAL04} who find no shifted bumps in a sample of over 400 Galactic 
sight lines. However, in this paper we will use the traditional 
six-parameter FM fits. 

In a different empirical analysis, Cardelli, 
Clayton \& Mathis (1988; 1989, hereafter CCM) used data from the infrared 
to the UV to find an average relationship for extinction as a function of 
wavenumber that relies on a single parameter, the ratio of 
total-to-selective extinction, $R_{V}$ = A$_{V}$/E(B-V). 
$R_{V}$ values are interpreted as being related to the grain environment
since they correlate with particle size in dust grain models. The
CCM relationship seems to be appropriate for the vast majority
($\sim 99\%$) of Galactic sight lines with known extinction curves
\citep{VAL04} indicating that extinction from the infrared (IR) to UV
varies in a systematic way among a wide range of dust environments.
A potential value of the CCM formulation is its general predictive
capabilities, providing an extinction curve from the IR through UV
(0.3 -- 8.7 $\mu$m$^{-1}$) 
based only on photometric observations at IR and optical wavelengths. 

The first measurements of extinction beyond 8.7 $\mu$m$^{-1}$ (below 1150 
\AA) were made with {\it Copernicus}. From these initial few sight lines: 
$\zeta$ Oph, $\xi$ Per, $\alpha$ Cam and $\sigma$ Sco \citep{YOR73, SY75} 
it became evident that the FUV curves are generally a continuation of the 
UV extinction trend, and that the extinction continues to rise out to 
10 $\mu$m$^{-1}$. CCM made a tentative extension of their extinction 
relation beyond 8 $\mu$m$^{-1}$  using the {\it Copernicus} data. 
Although the extrapolation was based only on a few sight lines, the 
congruence of these extinction curves demonstrated that the average 
$R_{V}$-dependent law extends beyond 8 $\mu$m$^{-1}$ in the Galaxy. 
In addition, CCM showed that a simple extension of the FM fit to FUV 
wavelengths gives, on average, a fair representation of the actual 
extinction in this spectral region. The \citet{SM79} ``average'' 
interstellar extinction curve beyond 9 $\mu$m$^{-1}$ is based entirely 
on {\it Copernicus} observations of three sight lines. In the nearly
thirty years since {\it Copernicus}, a few additional reddened sight lines 
in the Galaxy have been measured in the FUV using {\it Voyager}, {\it HUT}, 
{\it ORFEUS}, {\it FUSE} and suborbital rockets \citep{LON89, SAP90, 
GRE92, BUS94, HG01, SAS02, LCC05}. The extinction curves produced in these 
papers are broadly consistent with extrapolations of the FM and
CCM relationships into the FUV.

Though the cited studies revealed some evidence that extinction curves 
continue to rise beyond 8.7 $\mu$m$^{-1}$, the paucity of appropriate data
did not allow for a systematic study of FUV extinction, including a 
rigorous determination of whether an extrapolation of FM or CCM to FUV 
wavelengths would be appropriate in a diverse set of interstellar 
environments.
Instrumental issues of scattered light, 
time-variable sensitivity, and a limited sample of target and comparison 
stars have limited the utility of the {\it Copernicus} data set 
\citep{JEN86, SAP90}. The {\it Voyager} UVS data also suffer from low 
resolution 
which prevents explicit characterization and removal of the effects of
molecular hydrogen absorption. {\it Hopkins Ultraviolet Telescope} and
{\it ORFEUS} have produced insufficient
data for a  systematic survey, and rocket-borne instruments have not probed
diverse regions \citep{LCC05}. {\it FUSE} is the first instrument that
has produced an appropriate data set for undertaking a thorough and
careful analysis of FUV extinction over a wide range of sight line conditions.

This is the first in a series of three papers that will explore extinction in
the FUV. Paper II in the series \citep{CAR05} will present our first 
results for FUV extinction in the Magellanic Clouds. 
Paper III (in preparation)
will employ a much larger sample of sight lines ($\gtrsim$ 70) in order
to provide a greater statistical significance to trends in observed
extinction properties. The present study investigates FUV extinction 
characteristics in nine Galactic 
sight lines that span a wide range of dust environments as measured by $R_{V}$
(a proxy for dust grain properties). We explore how the extinction curves 
from 8.7 -- 9.5 $\mu$m$^{-1}$ (1150 - 1050 \AA) relate to 
extrapolations from the FM and CCM relationships. We also apply a simple 
model to the extinction curves in order to investigate how the addition
of FUV data might change the constraints on dust characteristics. Though our 
3-component grain model does not account for the full complexity
of dust in the interstellar medium, the model results do provide a rough
guide for estimating mass requirements among the grain components. 
In \S2 we present the observations, data reduction procedures, and modeling
algorithm. We discuss 
the nature of the sight lines, analyze the extinction curves in terms of 
FM and CCM fits, and consider model implications for the ``small grain'' 
population in \S3. A summary is given in \S4.

\section{Observations, Data Reduction and Analysis}

\subsection {Far Ultraviolet Spectra}
 
The FUV data used for this study were obtained with {\it FUSE} through the 
low-resolution (LWRS) aperture resulting in a resolution on the order of 
R = 20,000 and covering a wavelength range from approximately 950 to 1185 
\AA (or 10.5 -- 8.4 $\mu$m$^{-1}$).
The observations, originally obtained for a variety of programs
(including our own), were made
between October 1999 and September 2001.
The basic characteristics of the observations including the {\it FUSE}
target designations, the observing modes, and the exposure times are
summarized in Table 1. 
The data were retrieved from the {\it FUSE} archive and  
recalibrated using CALFUSE version 2.2.1. We found that calibrations 
with versions of CALFUSE earlier than 2.0 did not possess sufficient 
photometric accuracy for determining reliable extinction curves, for 
instance see \citet{HG01}. The eight calibrated spectra for each data set 
were mapped to a common wavelength
array spanning the entire range of the data. The data were merged by 
finding the mean flux at each wavelength point, weighted by three 
broad signal-to-noise bins determined for each individual
spectrum. The error arrays were merged in a manner consistent with that
of the data arrays.  For several observations, there were obvious
misalignments of the star in the aperture, resulting in low or absent flux 
values for individual spectra. Such bad data segments were flagged before
merging the spectra and were given zero weight in the co-addition. The
{\it FUSE} data anomaly known as the ``worm'' \citep{SAH02} was also
present in a majority of the observations. The flux values that
were lowered by this anomaly were also flagged, and thus did not contribute
to the final merged spectra. Merged spectra were co-added in the case
of multiple observations of an individual sight line. 

For our sample of stars, we note that the flux 
values in the wavelength overlap region between the {\it IUE} and {\it FUSE} 
data (1150 -- 1185 \AA) match well. This contrasts with the lower
signal-to-noise Magellanic Cloud sight line data of \citet{CAR05} who,
in some cases, needed to scale their {\it FUSE} data in order to provide
a better match to the {\it IUE} flux levels.

\subsection {Corrections for H$_2$ and H {\sc i}}

Absorption features produced by interstellar molecular hydrogen can have 
a substantial effect on the observed FUV spectrum of a star.  That is to
say, a considerable fraction of the light below $\sim$1108 \AA\ may be
absorbed by this molecule.  Although the
strongest lines of each H$_{2}$ band can reduce the observed flux to
nearly zero, the broad overlapping damping wings are able to be
modeled and removed (i.e., the spectrum is rectified to
the true continuum level).  In addition, the numerous narrow lines due to
rotationally excited states can be reliably removed, with the
exception of strongest line cores.  Even some of the lightly reddened
stars (some of our comparison stars) can show enough H$_2$ absorption that
we chose to remove it.  Our techniques for measuring the column densities
of each rotational state are described by \citet{RAC01,RAC02}.
In brief, we use profile fitting to measure the strongest lines,
and a curve of growth analysis for the weaker lines.  The latter
analysis also gives an ``effective'' $b$-value for the
rotationally excited lines, which is necessary because we
generally do not have velocity structure information for H$_2$.
We generate a model transmission spectrum based on the retrieved column
densities and $b$-value.  Because slight wavelength mismatches will
result in a poor removal of the H$_2$ spectrum, we perform
a running cross-correlation between the model and the observed
spectrum.  This produces a map of the appropriate wavelength
shift as a function of pixel number, and once those shifts are
applied to the model, one can simply divide the observed spectrum
by the model to remove H$_{2}$.  We flag (and exclude from further analysis)
all pixels for which the transmission model values fall below 30\% of the
continuum level. The spectra displayed in Figure 1 have the effects of 
H$_{2}$ removed where the spectrum is 70\% or more of the continuum level. 
The spectra are not plotted at wavelengths where the H$_{2}$ absorbs 
greater than 30\% of the continuum.

The solid vertical lines in Figure~1 indicate the locations of the 
H {\sc i} Lyman absorption features. Using a variant of the method
outlined by \citet{BOH75} (see Gordon et al. 2003 for details), 
we have estimated the column densities of H {\sc i} and have attempted to
reconstruct the continuum near the Lyman absorption.
As shown by the spectra in Figure 1, the Lyman features are not precisely
accounted for, particularly in the line cores.
We take a conservative approach and avoid the regions near the Lyman
lines when determining the extinction curves.
The spectral regions between the vertical dashed lines in Figure 1
show the regions around the Lyman features that have been
excluded from the extinction curve determinations below.  
The crowding of the Lyman lines and the profusion of H$_{2}$ absorption
at wavelengths shorter than Ly$\beta$ (1026
\AA\ or 9.75 $\mu$m$^{-1}$) means that there are no reliable 
continuum data above 9.5 $\mu$m$^{-1}$.

\subsection {Extinction Curves}

We combine data from 2MASS near-infrared photometry \citep{SKR97}, 
optical photometry (references are given in Table 1), archival low-dispersion
$IUE$ UV spectra processed with the \citet{MF00} method, and $FUSE$ FUV 
spectra for our target sight lines; the spectra were binned to 5 \AA.
We determined extinction curves from those data using the pair method 
\citep{MSF83}. The 
spectral matches between the reddened and comparison stars have been 
rigorously evaluated by examining their spectra in the IR through UV 
spectral regions. The stellar pairs used for this study are shown in 
Figure 1 where the spectrum of the reddened star in each pair lies below 
the comparison star's spectrum. The {\it FUSE} data, at wavenumbers 
greater than 8.4 $\mu$m$^{-1}$, obviously have lower signal-to-noise as 
compared to the longer wavelength data. Our method for assembling the
complete extinction curves from the IR to the FUV follows that of
\citet{GOR03} and \citet{GC98}. The derived 
extinction curves for each pair are shown in Figure 2. Note that the curves 
have a smooth transition from the {\it IUE} to {\it FUSE} wavelengths 
around 8.4 $\mu$m$^{-1}$.

\subsection {The Maximum Entropy Method Algorithm}

We employ the Maximum Entropy Method (MEM) algorithm for fitting the
extinction curves as developed by Kim, Martin \& Hendry (1994; hereafter 
KMH), with slight
modifications to facilitate application to the {\it FUSE} data.
Additional information regarding the MEM implementation may also be
found in \citet{HM00}.  As
recommended by KMH, we use the ``mass distribution'' -- m(a) da = mass
of dust grains in the grain-radius interval a to a+da -- instead of the more
traditional number of grains or ``size distribution.'' In this form,
the classical MRN-type model \citep{MRN77} becomes $m(a)
\propto a^{-0.5}$.
The grain sizes are divided into 50 logarithmically-spaced bins over
the range 0.0025-2.7 \micron.  As discussed by KMH, the shape of the
mass distribution is strongly constrained only for data over the
region $\sim$0.02-1 \micron.  Below a size of 0.02 \micron, the
extinction even in the ultraviolet becomes increasingly dominated by
absorption as sizes approach the small-particle (Rayleigh) limit and
thus constrains only total mass.  Above 1 \micron, the ``gray'' nature
of opacity (i.e., extinction efficiency approaching 2) provides a
similar type of integral constraint (on surface area).  Thus there is
danger of a large unconstrained mass if the default mass distribution
used in the definition of entropy is not well chosen.  We therefore
specify the defaults (templates) using the functional form of a
power-law with an exponential decay above some characteristic large
size $a_b$ (PED, see KMH): $m(a) \propto\ a^{-p} exp(-a/a_b)$.  
The PED is essentially the traditional gamma distribution function.

The ``observations'' to which the model is fit are taken from the FM
parameterization of the UV data and from the visible/infrared 
extinction data, resampled
at 34 wavelengths.  For model wavelengths longward of the observed data,
the extinction is calculated using the CCM relation.  However, their inclusion
in the modeling effort is an artifact of the current MEM implementation,
and we do not wish for them to contribute to the resulting mass distributions.
As a result, their errors are set to a large number that effectively reduces
their weight to zero.  The ``errors'' that are used to weight the data in
the MEM algorithm are calculated to represent only the scatter or random
errors in the extinction curve.  In the visible and infrared, these
values are taken directly from the random error estimate
produced during the extinction curve generation.  Unfortunately, the
same approach for the UV points tends to produce ``errors'' which are much
smaller than the point-to-point scatter in the data.  As a result, for each
model UV bin we have adopted the average deviation of the FM curve from 
the actual extinction data.

In this work, we consider (only) three-component models of
homogeneous, spherical grains: modified ``astronomical silicate"
\cite{wd01}, amorphous carbon \citep{zmcb96}, and graphite
\citep{ld93}, with mass densities ($\rho$) of 3.3 g/cm$^3$, 1.8 g/cm$^3$,
and 2.3 g/cm$^3$, respectively.  The graphite component is included, as
in many models of interstellar extinction, to produce the extinction bump at
2175~\AA.  Only small graphite particles are appropriate for this
purpose, and so for that component we chose $a_b = 0.02$~\micron, much
smaller than for the other components ($a_b = 0.25$~\micron).  At each
wavelength, the appropriately-weighted grain cross section is
integrated for each bin (and for each separate composition, see
equation 1 in KMH).  In order to incorporate abundance constraints
(see below), these are expressed as extinction per unit H column
density (accounting for H {\sc i} and H$_2$).

The total mass of dust is limited by the observed dust-to-gas ratio
and ``cosmic" abundances.
The MEM algorithm includes explicit constraints that do not allow the
model to use more carbon or silicon than is ``available''.  The cosmic
abundances adopted were 358 C and 35 Si atoms per million H.  The C
abundance is an estimate from young F and G stars \citep{SM01}, and
the Si abundance is solar \citep{HOL01}.  We note that the cosmic C
abundance used here is generous; recent results suggest that the solar
C/H may be as low as $245 \pm 23$ C per million H \citep{all02}.  In
order to allow generally for observed carbon gas-phase abundances, we
require that no more than 70\% of the adopted cosmic C may be used
(i.e., 251 C per million H), whereas we allow up to 120\% of the
cosmic Si (i.e., 42 Si per million H).
These relaxed constraints allow for uncertainties in abundances and
depletion, and in the dust-to-gas ratio (here $A_V/N[H_{tot}]$).  Note
that the model does not have to use this much C or Si, but often
does.

To provide consistency between the individual analyses, we use the
same abundance constraints, defaults, and initial guesses of the mass
distribtion (to which the model is not particularly sensitive).
In order to isolate the effect of including the {\it FUSE} data in the
analyses, each model is run with and without the {\it FUSE} data, (the
latter case giving no weight to data beyond 8.7 \micron$^{-1}$).
The MEM algorithm proceeds iteratively to find mass distributions
which reproduce the data.  The goodness of fit is judged by a
reduced-$\chi^{2}$, the ``target'' value of which should be close to
unity for an acceptable solution (we used 1.0 as our target).  In the
MEM algorithm used, this value is a constraint (along with abundance
limits, which must be satisfied for any solution to be considered valid.
In other words, a successful MEM solution will have a reduced-$\chi^{2}$
equal to the target value.  The assessment of
goodness of fit is relative to the adopted errors, and so it is
important that these be appropriately calculated (systematic errors
are not included). From this point of view, our assessment of random error
(see above) appears reasonable, producing ``good'' fits for a
reduced-$\chi^{2}=1$.

\section{Discussion}

\subsection{Sight line characteristics}

The basic characteristics of the nine sight lines used in this study are 
shown in Table 2. The reddened sight lines cover a wide range of Galactic 
environments as indicated by the large spread in $R_{V}$ values, from 2.43 
to 3.81 (see Table 2). These values of $R_{V}$ were determined using a 
$\chi^{2}$ minimization method; see \citet{GOR03} for the details of the 
method and associated uncertainties. As noted in the introduction, the 
extinction along these sight lines
is unusual compared to a typical line of sight measured in the local
interstellar medium.  Nevertheless, the sight lines toward HD 14250,
HD 73882, HD 99872, HD 167971, HD 239729, and HD 239683, though having
different values of $R_{V}$, are still fit well by CCM \citep{VAL04}. HD
14250 and HD 73882 were part of the original CCM sample.

HD 239729 and 239683 are in Trumpler 37 along with HD 204827 but do
not share any of the anomalous properties of the dust in the latter
sight line \citep{VAL03}.

HD 197770 shows a broad, weak spectral feature in the ultraviolet  
interstellar linear polarization centered close to 2175 \AA\ bump 
\citep{MAR95, WOL97, CLA92}. HD 197770 is an evolved, spectroscopic, 
eclipsing binary with two B2 stars \citep{GOR98}. 
The star seems to lie on the edge of a large area of molecular clouds and 
star formation including Lynds 1036 and 1049 in the Cygnus region 
\citep{GOR98}. On the {\it IRAS} maps there is a bright 60 \micron\ 
shell, 14\arcmin\ in radius centered on HD 197770, surrounded by an 
apparent bubble 
cleared of dust with a radius of approximately 24\arcmin\ 
\citep{GV93,GOR98}.

Only five Galactic sight lines (HD 29647, HD 62542, HD 204827, HD
210121, and HD 283809) are known to show systematic deviations 
from CCM, excluding the FUV \citep{VAL03, VAL04, cla03b}; these sight 
lines all sample dense, molecule-rich clouds.
Two of these sight lines, toward HD 62542 and HD 210121, both of which
show strong UV extinction, are included in our data set. The gas
toward HD 62542 is rich in CN and CH molecules \citep{car90}, and the
sight line lies on the edge of material swept up by a stellar wind
bubble.  The dust in the molecular cloud associated with HD 210121
\citep{LWH96, LAR00} seems likely to have been processed too as it was
propelled into the halo during a Galactic fountain or other event.
As for HD~62542, the non-CCM extinction curve toward HD 210121 has a
weak bump, and an even steeper FUV rise, indicating size distributions
that are skewed toward small grains \citep{VAL03,VAL04}, possibly from
exposure to shocks or strong UV radiation that disrupted large grains.

\subsection{FM and CCM Extrapolation to the FUV}

The empirical six-parameter FM relationship well reproduces Galactic 
UV extinction curves up to 8.7 $\mu$m$^{-1}$. CCM is able to reliably 
predict extinction curves up to 8.7 $\mu$m$^{-1}$ with the exception
of $\sim1\%$ of observed Galactic sight lines \citep{VAL04}. An extension of 
these relationships into the FUV would be valuable
for correcting extinction effects.  In addition, a valid extension of CCM
would be especially important because of its ability to predict (to some
degree) the extinction curves based on ground-based
photometric parameters alone.
Figures 3a and 3b show the extinction 
curves for our sample superimposed with three fits to each. The fits 
include each: an FM fit based on {\it IUE} data alone, an FM fit based on 
{\it IUE} and {\it FUSE} data, and a CCM fit based on the 
measured value of $R_{V}$ = A$_{V}$/E(B-V). 

Each of the Galactic extinction curves in our sample is well reproduced by 
FM curves fitted to data in the {\it IUE} and {\it FUSE} wavelength regions. 
The six parameters used for these FM fits are shown in Table 3. The good 
agreement between the extinction curves and FM fits in the FUV is not 
necessarily expected given that \citet{FM90} based their empirical fit 
to the extinction curves on data that went up to only 8.7 $\mu$m$^{-1}$ 
(i.e., only in the {\it IUE} wavelength region). \citet{FM88} found that 
the curvature of the extinction curve from 5.9 - 8.7  $\mu$m$^{-1}$ always 
has the same shape regardless of the dust size distribution or physical 
environment. Our small sample of sight lines suggest that this relationship 
extends to the FUV and that the extinction curve from 5.9 - 9.5 $\mu$m$^{-1}$ 
can be fit with a single continuous function; the same function found by 
\citet{FM88}. The continuity of the curvature shape at wavenumbers above 
8.7 $\mu$m$^{-1}$ is verified by the fact that the FM fits based on the 
{\it IUE} data alone are, in most cases, indistinguishable from the FM 
fits to the combined {\it IUE} and {\it FUSE} data (the solid and 
dotted lines in Figure 3). The largest deviation 
between the two FM fits for a given extinction curve occurs for HD 197770 
where the fits differ by less than 0.5 magnitudes at 9.5 $\mu$m$^{-1}$; 
this is 
within the expected uncertainty for this noisier portion of the extinction 
curve. Analysis of our sample suggests that FM fits can be extrapolated
safely from {\it IUE} wavelengths up to 9.5 $\mu$m$^{-1}$.
\citet{FM88} propose that the stability of the extinction curve shape
at short wavelengths results only from a dust component optical property,
as opposed to a function involving grain size.  While we certainly find
a similar behavior in shape of the extinction curve from
1050 - 1700 \AA, we cannot exclude the possibility of systematic variations
in particle sizes as viable mechanisms.

For these lines of sight, selected for their unusual extinction, the  
predictive capability of the CCM relationship for FUV extinction 
is more problematic. Note, however, that in contrast to the FM fits, no 
ultraviolet data, neither UV nor FUV, are used to determine the single
CCM fitting paramater $R_{V}$; thus the ability to predict the UV 
extinction for most stars is a remarkable feature of CCM. 
For the two sight lines in our sample where CCM does
not match even the UV extinction, toward HD 62542 and HD 210121, it also
fails to reproduce the FUV. This is not a surprising result since the
longer wavelength data already suggested that the grains in these sight
lines have likely experienced processing that is different from that in
more ``average'' Galactic 
sight lines. The departure from the CCM 
relationship grows with increased wavenumber over the measured region up 
to a maximum of about 3 magnitudes per A$_{V}$ at 9.5 $\mu$m$^{-1}$; for
HD 62542 and HD 210121, CCM underpredicts the FUV extinction as much as
40\% and 35\%, respectively. HD 62542 is known to have an anomalous 
2175\AA\ bump -- unusually weak and peaking at an atypically high 
wavenumber; \citet{cs88} find $x_{0} = 4.74 \micron^{-1}$ and we find
$x_{0} = 4.78 \pm 0.08 \micron^{-1}$. This suggests that the failure of CCM 
to reproduce the FUV extinction may be associated with anomalous properties
of the particles producing the 2175 \AA\ bump along this sight line.

The sight lines toward four stars, HD 73882, HD 99872, HD 167971 and HD 
239729, are, within uncertainties, well represented by CCM over 
the IR to UV wavelength region, but much less so in the FUV. The CCM
functions for these sight lines differ from 
the well-fit FM curves by 2 $\sigma$ or more (based on the $R_{V}$ 
uncertainties) at 9.5 $\mu$m$^{-1}$. As with the HD 62542 and HD 210121 
extinction curves, the differences between the CCM predictions and the 
measurements increase with increasing wavenumber. The fact 
that CCM fits the extinction curves in the IR through UV, but diverges in 
the FUV may indicate that the systematic processing of larger grains
does not necessarily continue to the small grain population, at least in
the same way. CCM FUV extinction in two of these sight
lines (HD 73882 and HD 239729) is too high 
while it is too low in the other two (HD 99872 and HD 167971). The $R_{V}$ 
values for this group of four sight lines ranges from 2.99 to 3.81.

The final three sight lines in our sample, toward HD 14250, HD 197770 and 
HD 239683, have CCM curves that predict well the FUV extinction.
The $R_{V}$ values of these three sight lines range from 
2.44 to 2.98; smaller than the group for which CCM did poorly.
While one might suggest that CCM can be extrapolated into the FUV more
reliably for smaller $R_V$ values and the potential connection to an
increased relative abundance of small grains, further generalization
awaits a larger sample of FUV extinction (as will be presented in
Paper III).

\subsection {MEM Grain Models}

The resulting model-data comparisons are presented in Figure 4, which
shows the amount of extinction
provided by each of the three materials and the total extinction of
the model compared to the measured extinction curve. The percentages
of the carbon and silicon along the sight lines that are used to
produce the displayed fits are given in Table 4.  

Our models use three grain types to reproduce the observed extinction, 
specifically astronomical silicates, amorphous carbon, and graphite.
A more realistic dust model would include other compositions such as oxides 
and grains composed of a combination of carbonaceous and silicaceous 
materials. It would also account for non-spherical particles and the effects 
of coatings on and vacuums in the grains. We have chosen to use a simple 
model here because 
of the computational complexity and arbitrary assumptions often associated 
with a more physically complete approach \citep{cla03}. 
By doing so, we acknowledge that the detailed results of the models, 
such as specific populations of grain types and sizes, and the 
abundance constraints that they imply, are to be used with caution. However, 
these simple models can provide some useful information about dust. Since 
the absorption characteristics of a grain may be well constrained by its bulk 
behavior, the models can provide estimates of the mass distribution among 
grain-size populations that are appropriate to more complex systems.

The left and right columns in Figure 4 show the MEM fits to the extinction 
curves based on data up to 9.5 $\mu$m$^{-1}$ 
({\it FUSE} wavelengths), and 8.7 $\mu$m$^{-1}$ ({\it IUE} wavelengths), 
respectively. 
Although some of the FUV extinction 
is produced by the grains that are also responsible for extinction at longer 
wavelengths, it is clear that models based on data only to 8.7 $\mu$m$^{-1}$ 
do not always fit the FUV extinction well. Other than toward HD 167971 
and HD 197770, the extinction curves require a larger fraction of grain 
material to be in small
dust particles, mostly silicates in our model, to adequately account for the
FUV. More specifically, within the context of our simple three-component
model, the inclusion of the FUV requires up to 10\% more silicon to be 
in grains (see Table 4). Since the FM parameter $c_{4}$ is a measure of the
strength of the far UV rise in an extinction curve, it is not surprising that
we find a correlation between this value and an increased need for small 
silicate grains. A linear fit to this 
relationship (a higher-order polynomial is not statistically justifiable) 
indicates that in order to fit the FUV constraint, 12\% more silicon is 
needed per increase of 1.0 in $c_{4}$; the $1 \sigma$ scatter about the 
line is a 2\% change in the
required silicon. The largest portion of the scatter is likely attributable to
the errors assigned to the extinction curve data (see \S2.4). 
Since the models tend to underfit the extinction when less 
constrained (when the data have larger uncertainties; see the left panels in 
Figure 4), we are likely underestimating the 
extra mass needed in small silicates for our simple model to fit the FUV.
The fraction of C allocated to amorphous carbon grains often decreases 
in the models that include the FUV extinction
constraint (see Table 4); amorphous carbon serves as a source of visible 
opacity in these models. On the other hand, the abundance of C that the 
models allocate to graphite grains is often enhanced when the FUV
is included.  This general increase in graphite and 
decrease in amorphous carbon represents the model's reallocation of C to 
fit better the extinction out to 9.5 \micron$^{-1}$; the upward curvature 
in the graphite extinction at wavenumbers greater than 8 $\mu$m$^{-1}$ fits 
the upward turn often seen in the extinction curves at these wavelengths. 
Nevertheless, one should also consider the possibility that there is
something wrong with the dielectric function of our silicate component
in not being able to produce this upward curvature, or perhaps that
other components missing entirely, e.g., PAHs, nanoparticles.

We note that the fits to x $< 9.5$ \micron$^{-1}$ in Figure 4 are 
often poorest in the FUV where 
the models tend to underestimate the extinction. Since a disproportionate 
share of the target $\chi^{2}$ is coming from the FUV, we have 
experimented with lowering the target reduced-$\chi^{2}$ in the models 
to better fit this region. We show two examples of forcing a tighter FUV
fit in Figure 5. The figure displays three MEM models for the sight lines 
toward HD 62542 and HD 210121, one based on each: data up to 8.7
\micron$^{-1}$ with a reduced-$\chi^{2}=1.0$, data up to 9.5 
\micron$^{-1}$ with a reduced-$\chi^{2}=1.0$, and data up to 9.5 
\micron$^{-1}$ with a reduced-$\chi^{2}=0.5$. 
The left panels show the fits to the data for each model. The
right panels show the resultant mass distribution of each component relative
to the mass of hydrogen.
The reduced-$\chi^{2}=1$, 
x $<9.5$ \micron$^{-1}$ model for HD 62542 does not use all of the Si or C 
available to it, so the model can draw on both of these elements when it
tries to fit better the FUV data (see Table 4). The 
reduced-$\chi^{2}=0.5$ model does, in fact, take the entire allocation of
Si (120\%), however it uses less of the available C than the less-constrained
fit; specifically, the more-constrained model puts 30\% of the total carbon 
abundance into amorphous C and 7\% into graphite. 
The top right panel of Figure 5 shows that much of 
the ``extra'' silicon used to fit better the FUV was 
put into small grains. The grain sizes below 200 \AA\ are not 
well resolved, so the shape of the mass curve at these small sizes is guided 
by the model template and not the actual distribution in size.  However, 
the changes in the overall mass of grains below 0.02 $\mu$m, as shown in 
the figure, are significant. Figure 5 also shows that the model is
reallocating the size distributions of the grains so that, for instance, 
the graphite is contributing more extinction at 2175 \AA\ relative to 
the amorphous carbon (note the decrease in amorphous carbon grains
with sizes around 0.2 \micron\ as the FUV is better fit). 
Recall however that at reduced-$\chi^{2}=0.5$, the model is 
overconstrained if our treatment of the errors is appropriate. 
We caution against overinterpreting these size distribution
results; we present them merely to show that this simplified model requires
changes in the grain size distributions in order to better fit the FUV.

The reduced-$\chi^{2}=1$ model for HD 210121 uses the maximum available Si, 
so it can only draw on C in order to put more mass into grains for a 
better FUV fit. The reduced-$\chi^{2}=0.5$ model 
continues to incorporate the maximum allocated Si, and increases 
the C that it uses for each of the amorphous carbon grains and graphite 
by 1\%  of the total C abundance. This overconstrained model is able
to fit the extinction primarily by reallocating the size-distributions
of the grains in a way that is similar to the reduced-$\chi^{2}=0.5$ model 
for HD 62542. A rigorous analysis that relates grain size distributions to
extinction will require a more complex model as well as
extinction curves with smaller uncertainties in the FUV.

\section{Summary}

1. We present the extinction curves from the IR to the FUV (up to 9.5 
$\mu$m$^{-1}$) for nine Galactic sight lines observed with {\it FUSE}.

2. The FUV extinction curves out to 9.5 $\mu$m$^{-1}$ are all well fit 
with an extrapolation of the FM relationship \citep{FM90}
from the {\it IUE} wavelength 
range that extends to only 8.7 $\mu$m$^{-1}$.

3. The CCM relationship \citep{ccm89} does not properly predict 
the FUV extinction in the two sight lines in our sample where it also fails 
to reproduce the IR -- UV extinction. 

4. CCM does fit the IR -- UV in seven of our sight lines. For three of these, 
all with $R_{V}$ values below the Galactic average, an extrapolation of CCM 
does properly predict the FUV extinction. The other four sight lines all have 
larger $R_{V}$ values than the previous group. An extrapolation of CCM 
over-predicts the FUV extinction in two of these sight lines, and 
under-predicts it in the other two.

5. Our simple MEM grain models are generally able to fit the extinction 
curves that include the FUV constraints. The more highly constrained
models usually require more mass to be in small grains (radii $\lesssim 
200$ \AA) 
in order to reproduce an extinction curve that includes the IR -- FUV as 
compared to those using only the IR -- UV.

\acknowledgments
This work was supported by the NASA grants NAG5-108185, NAG5-9249 and 
NAG5-7993.

\clearpage





\begin{figure}

\plotone{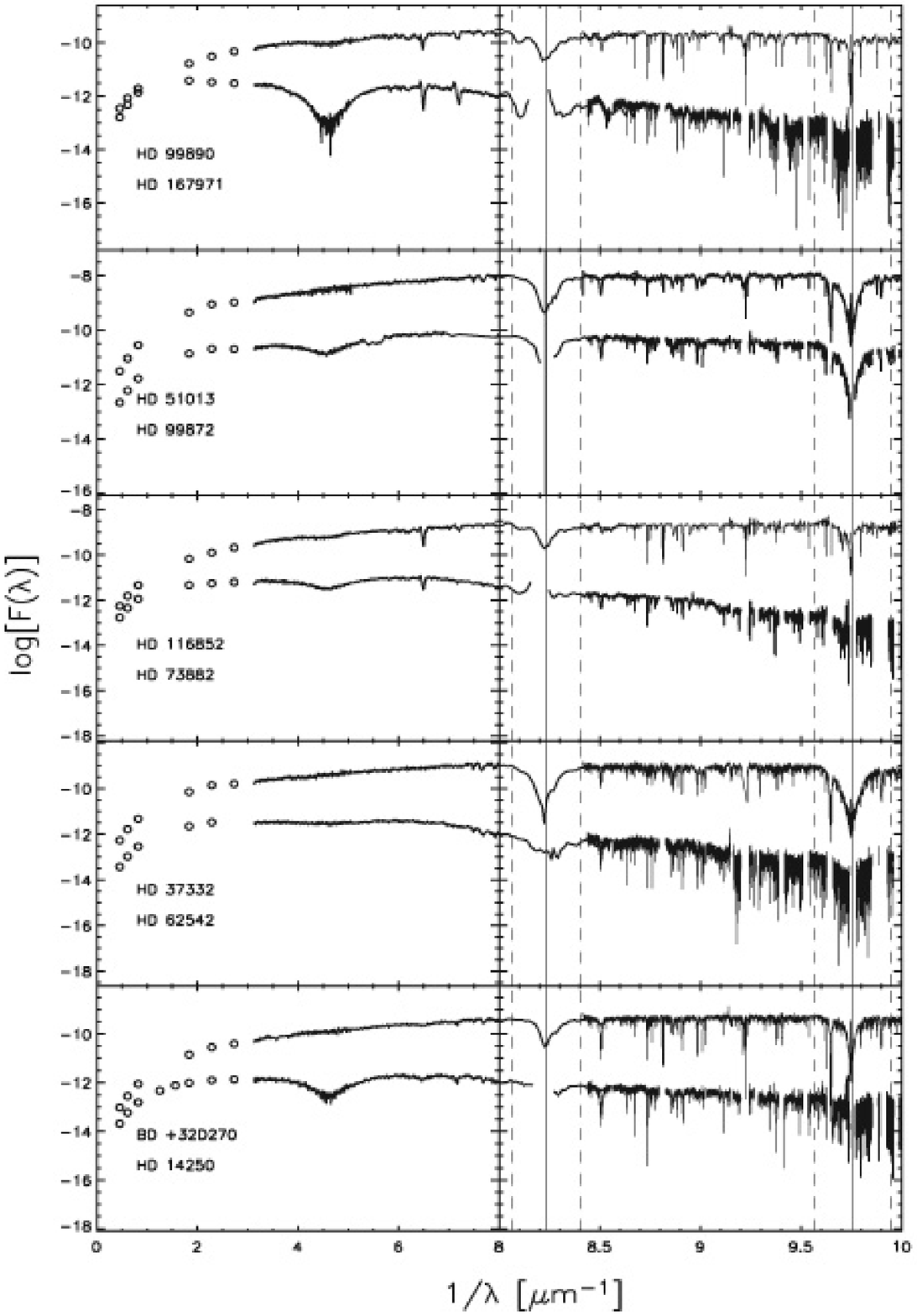}
\figurenum{1}
\caption{The infrared to far ultraviolet spectra of our sample pairs. The 
comparison and reddened stars are the upper and lower spectra in each panel,
respectively. The solid vertical lines show the locations 
of H {\sc i} absorption and the dashed vertical lines show the width 
of those features. The effects of H$_{2}$ absorption have been removed from the
spectrum where the absorption is less than or equal to
30\% of the continuum level. The
spectra are not plotted where the H$_{2}$ features absorb more than 30\%
of the continuum.}
\end{figure}

\newpage

\begin{figure}

\plotone{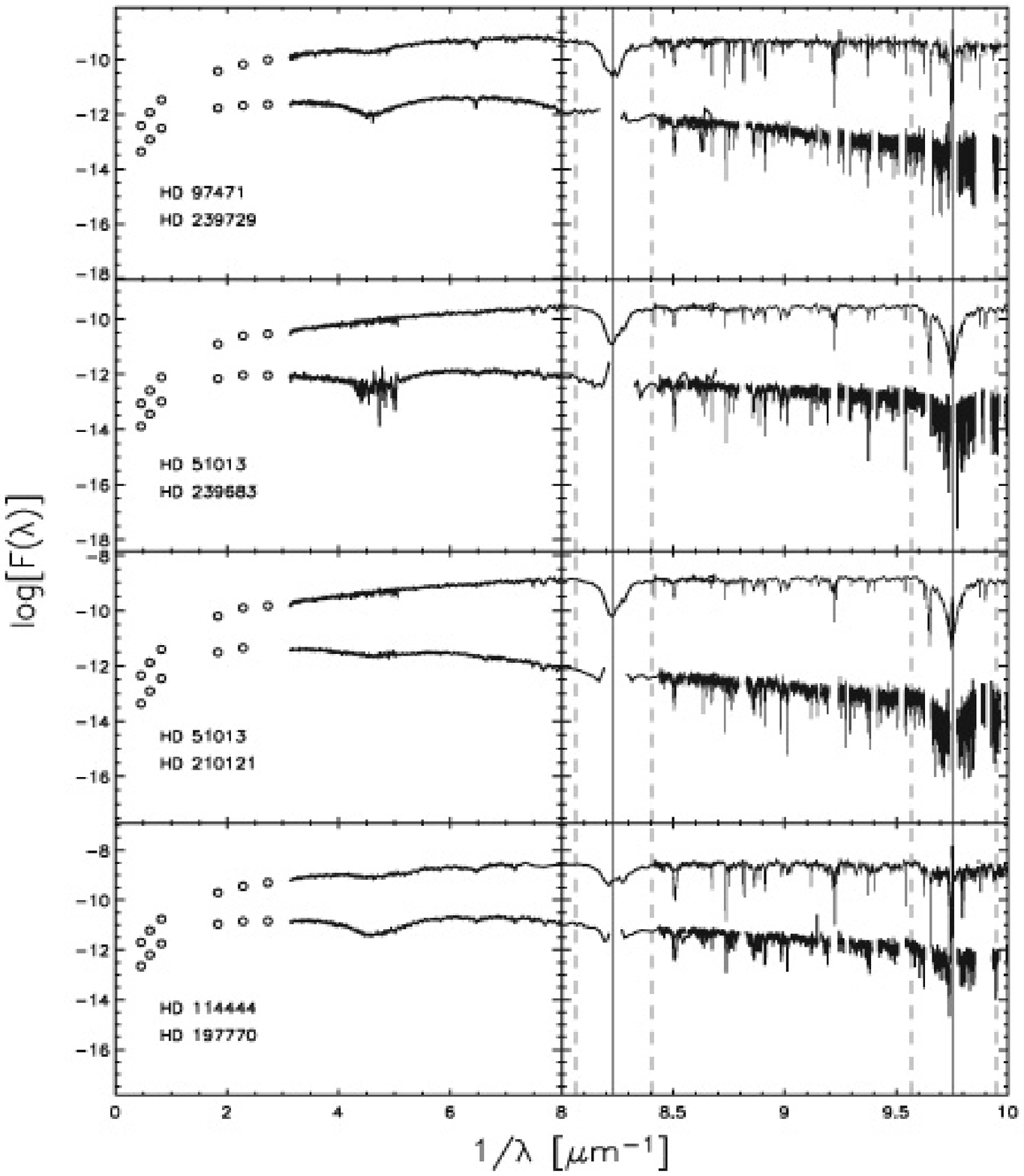}
\figurenum{1b}
\caption{The same as Figure 1a.}
\end{figure}

\newpage

\begin{figure}
\plotone{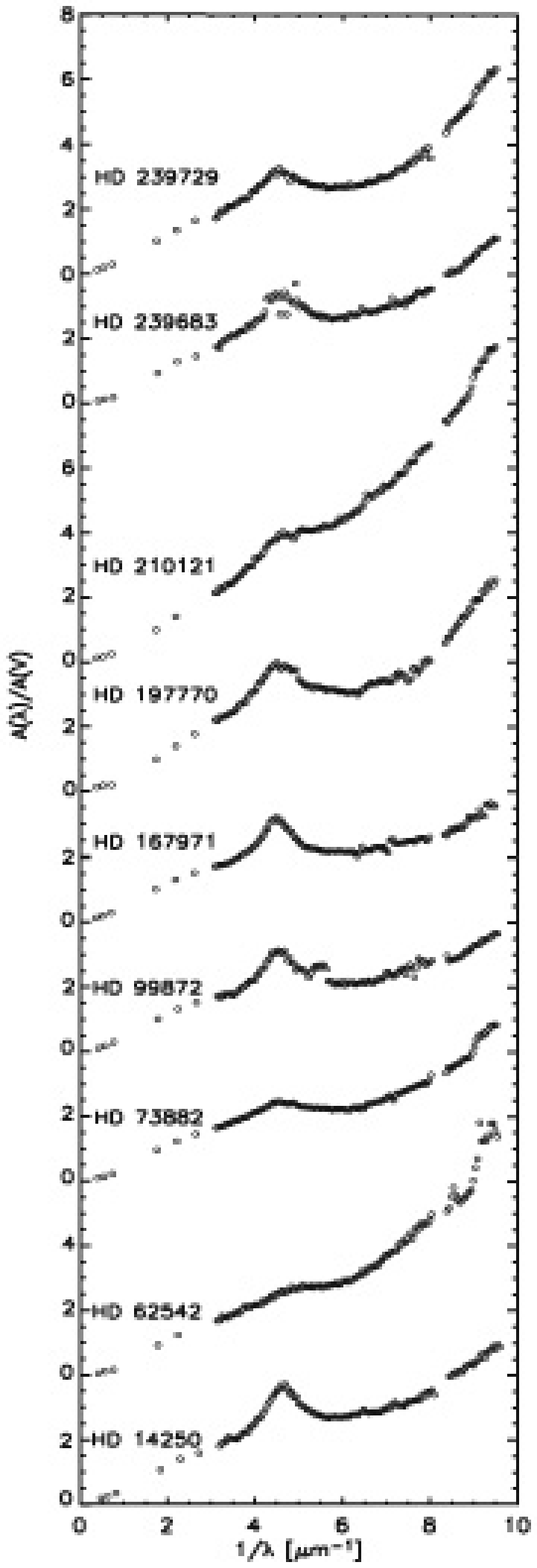}
\figurenum{2}
\caption{Extinction curves based on the spectral pairs in Figure 1.}
\end{figure}

\newpage

\begin{figure}
\figurenum{3}
\plotone{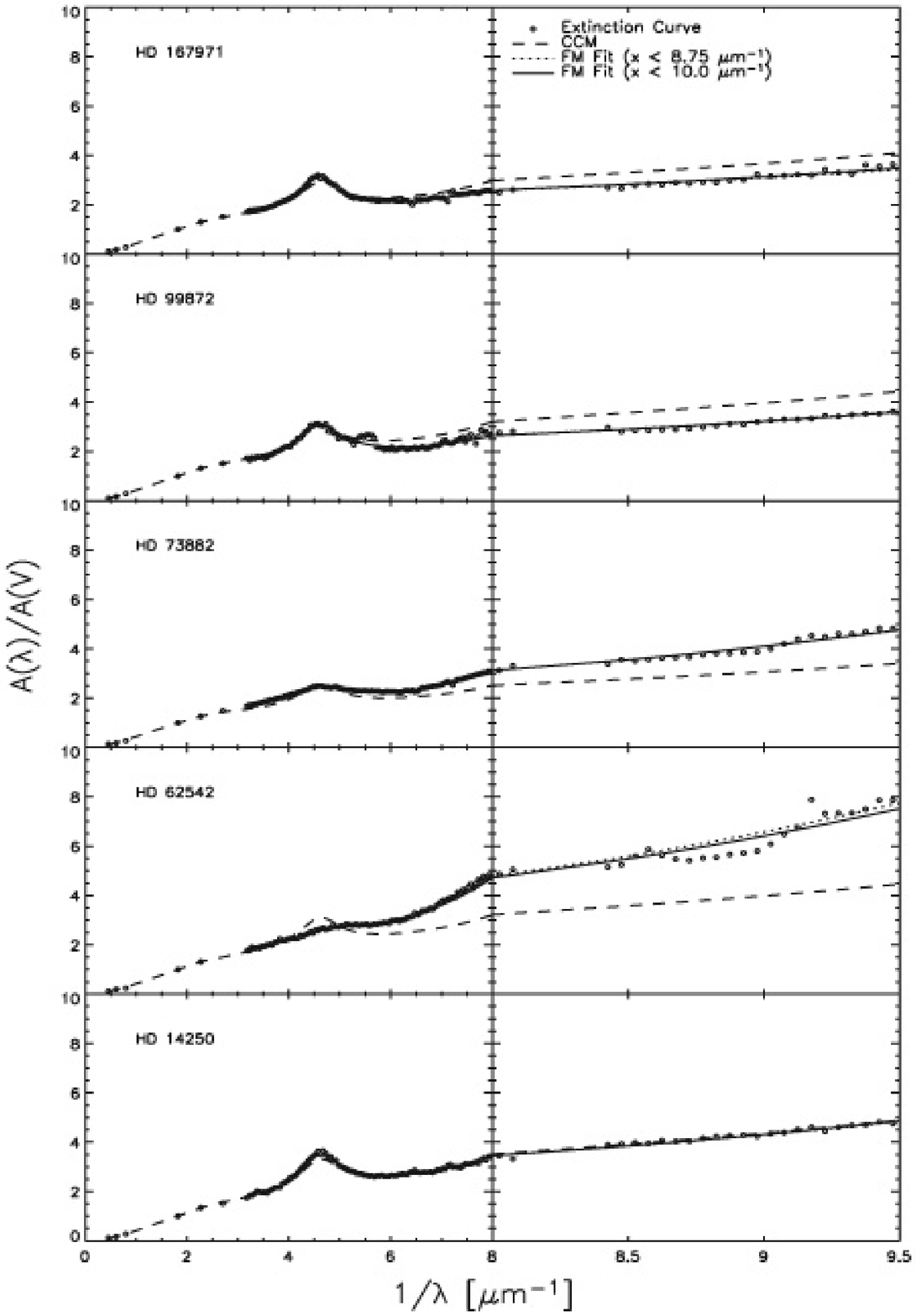}
\caption{The figure shows CCM and FM fits to the extinction curves; the
circles represent the measured extinction. The solid line that represents
the FM curves based on fits out to 9.5\micron$^{-1}$ are difficult to see 
because they so closely follow the extinction curves. 
The dotted line that represents the FM fit based only 
on data below 8.7 \micron$^{-1}$ is usually indistinguishable from the 
FM fit out to 9.5 \micron$^{-1}$. The dashed lines represent CCM curves based 
on the measured $R_{V}$ values along the sight lines (see Table2).}
\end{figure}

\newpage

\begin{figure}

\plotone{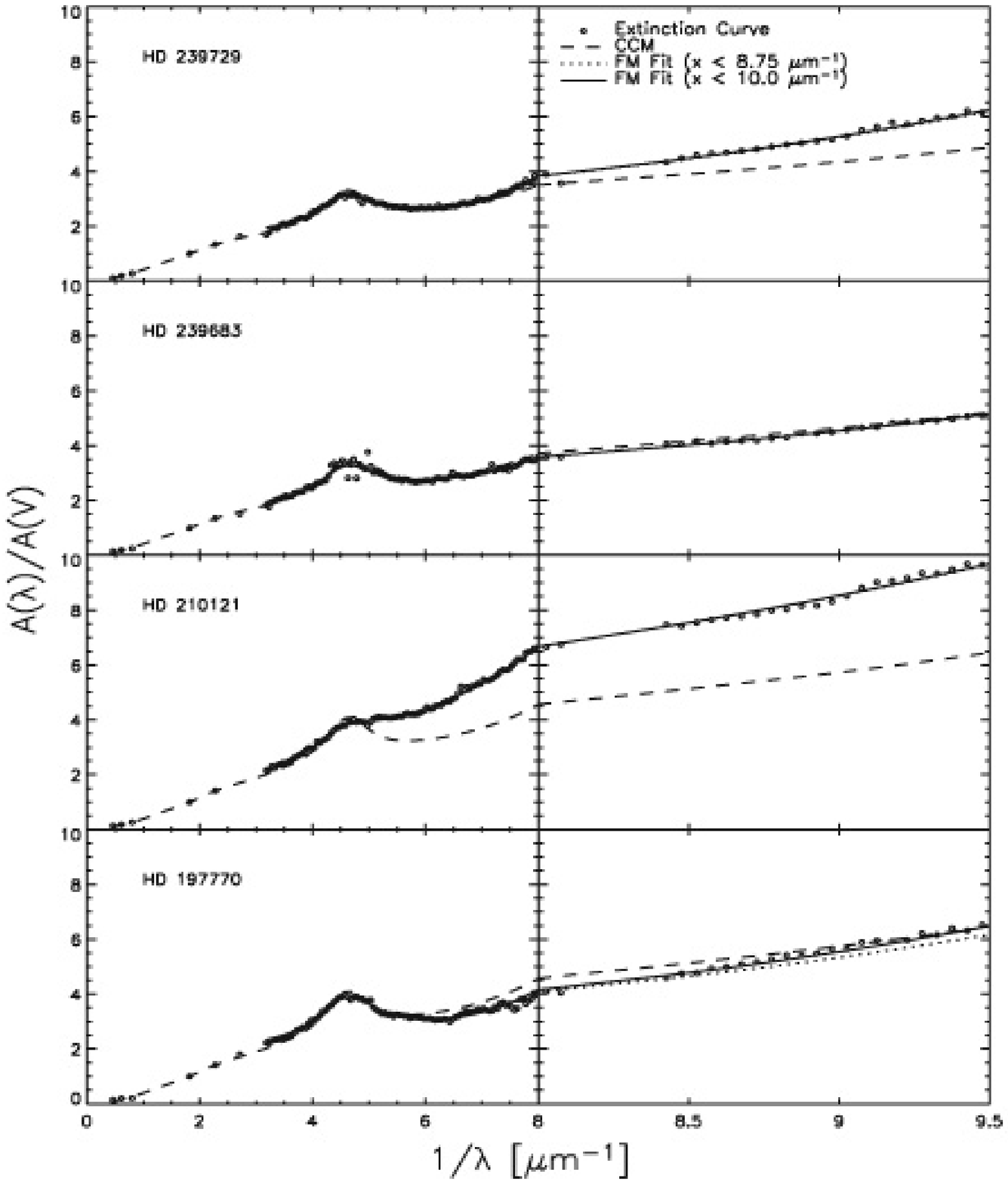}
\figurenum{3b}
\caption{The same as Figure 3a.}
\end{figure}

\newpage

\begin{figure}
\figurenum{4}
\plotone{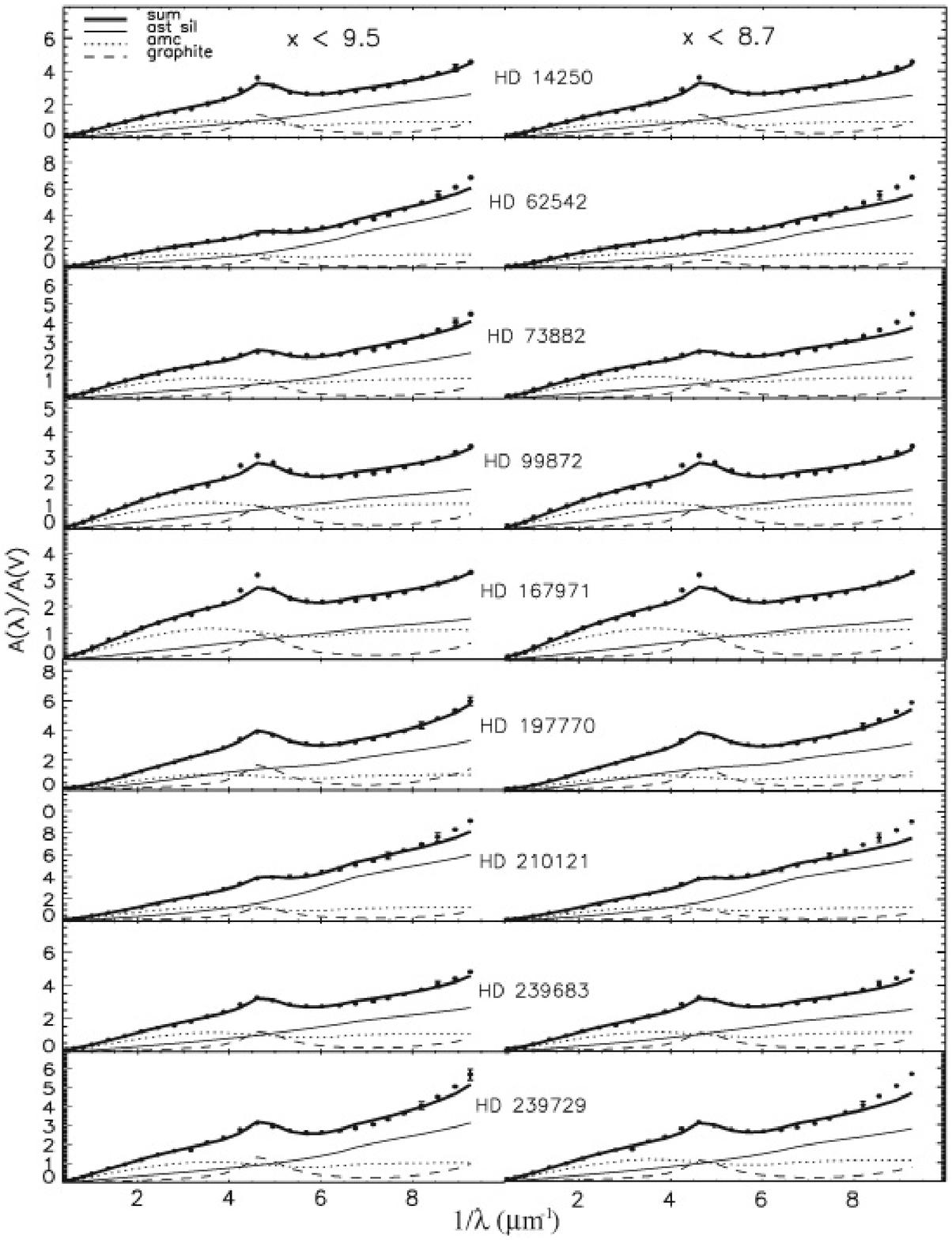}
\caption{The extinction curve fits resulting from the MEM modeling.
The models that fit out to the FUV (x $< 9.5$ \micron$^{-1}$) are shown in
the left panel, and those fit out to only the UV (x $< 8.7$ \micron$^{-1}$)
are on the right. The 1-$\sigma$ error bars (see \S4.1) are shown on every 
third point when they are larger than the point itself.
The circles are the FM fits to the data out to 9.5 \micron$^{-1}$ (see
Table 3). The thick solid line is the total extinction from three components:
astronomical silicate (narrow solid line), amorphous carbon (dotted line) and 
graphite (dashed line) grains. The fractions of silicon and carbon that 
the model incorporates into the different grain types are listed in Table 4.}
\end{figure}

\clearpage

\begin{figure}
\figurenum{5}
\plotone{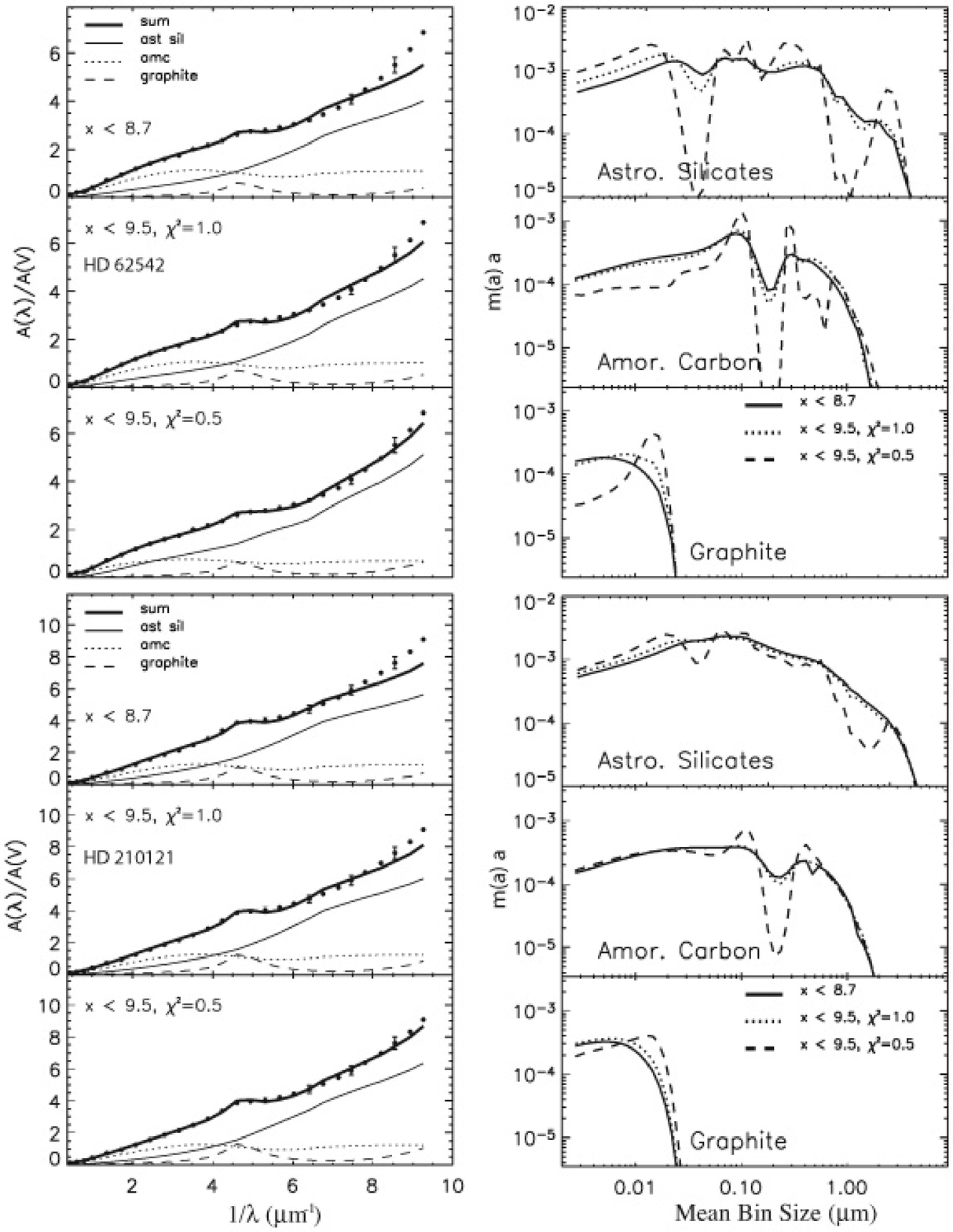}
\caption{MEM extinction fits (left panels) and mass distributions 
as a function of grain size (right panels)
for the sight lines toward HD 62542 (upper panels) and HD 210121
(lower panels). MEM models are shown for fits out to the UV data only
(x $< 8.7$ \micron$^{-1}$) with a reduced-$\chi^{2}=1.0$, for fits out to
the FUV (x $< 9.5$ \micron$^{-1}$) with a reduced-$\chi^{2}=1.0$, and for
fits out to the FUV with a reduced-$\chi^{2}=0.5$. The panels on the right 
show the distribution of grain sizes found for the three 
dust types modeled.}
\end{figure}

\clearpage
\begin{deluxetable}{llllc} 
\tablecolumns{5}
\tablewidth{0pc} 
\tablecaption{Obervations}
\label{table_obs}
\tablehead{
\colhead{Sight line} & \colhead{FUSE} &
\colhead{Exposure} & \colhead{Mode\tablenotemark{a}} & \colhead{Optical data} \\
\colhead{} & \colhead{targets} & \colhead{(sec)} & 
& \colhead{references\tablenotemark{b}}\\
}
\startdata 
BD+32 270&A063-07& 4,900&HIST & 1\\
HD 14250 &A118-06& 8,200&TTAG & 2\\
HD 37332 &B060-12& 2,300&HIST & 3\\
HD 51013 &A063-09& 4,100&HIST & 4\\
HD 62542 &P116-02&11,400&TTAG & 5\\
HD 73882 &X021-03&25,500&TTAG & 6\\
         &P116-13&13,600&TTAG & \\
HD 97471 &A118-04& 4,200&HIST & 7\\
HD 99872 &A120-06& 3,300&HIST & 5\\
HD 99890 &P102-46& 4,600&HIST & 7\\
HD 114444&A118-05& 5,500&TTAG & 8\\
HD 116852&P101-38& 5,100&HIST & 8\\
HD 167971&P116-21& 9,500&TTAG & 9\\
HD 197770&A118-13& 6,000&TTAG & 10\\
HD 210121&P116-30&13,800&TTAG & 11\\
HD 239683&A118-10& 9,200&TTAG & 12\\
HD 239729&A118-07& 9,900&TTAG & 12\\
\enddata 

\tablenotetext{a}{HIST and TTAG refer to the histogram and time tag
(photon list) data saving modes of {\it FUSE} \citep{SAH96}.}
\tablenotetext{b}{(1) \citet{DWC82}, (2) \citet{MEN67}, (3) \citet{HHT64},
(4) \citet{FEI67}, (5) \citet{CS62}, (6) \citet{DEN77}, (7) \citet{FEI69},
(8) \citet{HKV74}, (9) \citet{FOR84}, (10) \citet{HL77}, (11) \citet{WF92},
(12) \citet{GK76}.}
\end{deluxetable}

\newpage

%
\begin{deluxetable}{lllcclccccc} 
\tabletypesize{\footnotesize}
\tablecolumns{11}
\rotate 
\tablewidth{0pc} 
\tablecaption{Observational Characteristics of Sight Lines.
\label{table_characteristics}}
\tablehead{
\colhead{} & \colhead{} & \multicolumn{2}{c}{Reddened}
& \colhead{} & \multicolumn{2}{c}{Comparison}\\
\cline{3-4} \cline{6-7}
\colhead{Reddened} & \colhead{Comparison} &
\colhead{Sp T} & \colhead{$N_{H_2}$$^{a}$} & \colhead{} &
\colhead{Sp T} & \colhead{$N_{H_2}$} &
\colhead{R$_V$} & \colhead{$\Delta$(B-V)} & \colhead{$\Delta N_{HI}$} &
\colhead{H {\sc i} References$^b$} \\
\colhead{sight line} & \colhead{sight line} & \colhead{} &
\colhead{($10^{20} cm^{-2}$)} & \colhead{} &
\colhead{} & \colhead{($10^{20} cm^{-2}$)} & \colhead{} &
\colhead{(mag)} & \colhead{($10^{21} cm^{-2}$)}
}
\startdata 
HD 14250 &BD+32 270&B1 IV & 5.5&&B2 V  &\nodata&
2.98$\pm$0.14&0.48$\pm0.05$&1.57$\pm$0.28& 1\\
HD 62542 &HD 37332 &B5 V  & 6.5&&B5 V  &\nodata&
3.18$\pm$0.22&0.31$\pm$0.03&1.53$\pm$0.17& 2\\
HD 73882 &HD 116852&O9 III&12.9&&O9 III&0.62   &
3.81$\pm$0.16&0.49$\pm$0.02&1.29$\pm$0.53& 3\\
HD 99872 &HD 51013 &B3 V  & 3.5&&B3 V  &\nodata&
3.20$\pm$0.22&0.31$\pm$0.03&0.60$\pm$0.15& 1\\
HD 167971&HD 99890 &B0 V  & 7.1&&B0.5 V&\nodata&
3.37$\pm$0.09&0.81$\pm$0.03&3.99$\pm$0.27& 2,4\\
HD 197770&HD 114444&B2 III&10.7&&B2 III&\nodata&
2.44$\pm$0.15&0.37$\pm$0.02&0.63$\pm$0.05& 1\\
HD 210121&HD 51013 &B3 V  & 5.6&&B3 V  &\nodata&
2.43$\pm$0.16&0.35$\pm$0.05&1.41$\pm$0.32& 1\\
HD 239683&HD 51013 &B3 IV & 5.6&&B3 V  &\nodata&
2.86$\pm$0.15&0.43$\pm$0.03&2.00$\pm$0.47& 1\\
HD 239729&HD 97471 &B0 V  &11.8&&B0 V  &0.80   &
2.99$\pm$0.18&0.37$\pm$0.04&1.82$\pm$0.22& 2
\enddata 

\tablenotetext{a}{HD 62542, 73882, 167971, 210121 from Rachford
et al. (2002).}
\tablenotetext{b}{ (1) Ly-$\alpha$ fitting, this paper.
(2) N(H {\sc i})=4.9($\pm$0.23) $\times 10^{21}$ E$_{B-V}$
(Diplas \& Savage 1994). (3) Fitzpatrick \& Massa (1990). (4) Rachford 
et al. (2002) give 4.0($_{-2.0}^{+4.0}) \times 10^{21}$ cm$^{-2}$, which
is consistent with the application of (2). }
\end{deluxetable}

\clearpage

%
%
\begin{deluxetable}{lcccccc} 
\tablecolumns{7}
\tablewidth{0pc} 
\tablecaption{Parameters for Fitzpatrick \& Massa Fits to the Extinction 
Curves\tablenotemark{a}}
\label{table_fm}
\tablehead{
\colhead{Sight line}& \colhead{c$_1$} &  \colhead{c$_2$} &  \colhead{c$_3$} &
\colhead{c$_4$} &  \colhead{x$_0$} &  \colhead{$\gamma$}\\
}
\startdata 
\underline{HD 14250}\\
 x $<$ 9.5 \micron$^{-1}$  &   -0.351 &  0.779 &   3.945 &     0.454 &    4.593 &     0.941\\
 error   &     1.031 &     0.082 & 0.507 &     0.044 &     0.010 &     0.022\\
 x $<$ 8.7 \micron$^{-1}$ & -0.352 & 0.780 &     3.942 &     0.448 &    4.593 &     0.941\\
 error   &  0.079 &   0.076 &     0.451 &     0.052 &     0.012 &     0.016\\
\underline{HD 62542}\\
 x $<$ 9.5 \micron$^{-1}$ &    -1.240 & 1.253 &   0.446 &     1.047 &     4.780 &     0.798\\
 error   &     0.395 &     0.141 & 0.124 &     0.128 &     0.080 &     0.014\\
 x $<$ 8.7 \micron$^{-1}$ &    -1.203 & 1.242 & 0.486 &     1.131 &     4.787 &     0.826\\
 error   &     0.401 &     0.146 & 0.137 &     0.158 &     0.080 &     0.014\\
\underline{HD 73882}\\
 x $<$ 9.5 \micron$^{-1}$ &     0.879 & 0.593 &  3.123 &     0.810 &     4.603 &     1.248\\
 error   &     0.219 &   0.046 &  0.266 &     0.046 &     0.030 &     0.021\\
 x $<$ 8.7 \micron$^{-1}$ &     0.817 & 0.604 &  3.168 &     0.800 &     4.600 &     1.254\\
 error   &     0.163 &  0.040 &  0.234 &     0.049 &     0.020 &     0.020\\
\underline{HD 99872}\\
 x $<$ 9.5 \micron$^{-1}$ &  0.101 &  0.488 &     5.830 &     0.360 &     4.590 &     1.180\\
 error   &     0.038 &  0.080 &  1.310 &     0.097 &     0.076 &     0.020\\
 x $<$ 8.7 \micron$^{-1}$ &     0.097 & 0.487 & 5.862 &     0.370 &     4.590 &     1.182\\
 error   &     0.033 &     0.081 & 1.312 &     0.115 &     0.076 &     0.020\\
\underline{HD 167971}\\
 x $<$ 9.5 \micron$^{-1}$  &     1.538 &  0.330 &  2.811 &     0.381 &     4.577 &     0.808\\
 error   &     0.256 & 0.050 &     0.288 &     0.038 &     0.021 &     0.013\\
 x $<$ 8.7 \micron$^{-1}$  &     1.450 & 0.347 & 2.826 &     0.351 &     4.575 &     0.810\\
 error   &     0.249 &     0.044 & 0.277 &     0.044 &     0.018 &     0.013\\
\underline{HD 197770}\\
 x $<$ 9.5 \micron$^{-1}$ &     1.217 &     0.517 & 5.359 &  0.733 &     4.616 &     1.233\\
 error   &     0.326 &     0.056 &     0.586 & 0.058 &     0.032 &     0.021\\
 x $<$ 8.7 \micron$^{-1}$ &     1.112 &     0.546 & 5.115 & 0.637 &     4.609 &     1.207\\
 error   &     0.298 &     0.053 &     0.496 & 0.083 &     0.030 &     0.020\\
\underline{HD 210121}\\
 x $<$ 9.5 \micron$^{-1}$ &    -3.296 &     1.882 &  1.913 & 0.672 &     4.611 &     1.093\\
 error   &     0.676 &     0.262 & 0.450 &     0.106 &     0.066 &     0.018\\
 x $<$ 8.7 \micron$^{-1}$ &    -3.314 & 1.887 & 1.902 &     0.663 &     4.609 &     1.090\\
 error   &     0.760 &     0.268 & 0.454 &     0.119 &     0.064 &     0.018\\
\underline{HD 239683}\\
 x $<$ 9.5 \micron$^{-1}$ &     0.116 &     0.703 & 4.726 &   0.515 &     4.614 &     1.235\\
 error   &     0.031 &     0.057 &     0.601 & 0.047 &     0.044 &     0.022\\
 x $<$ 8.7 \micron$^{-1}$ &     0.137 &     0.699 & 4.738 & 0.523 &     4.616 &     1.237\\
 error   &     0.077 &     0.050 & 0.560 &     0.068 &     0.046 &     0.025\\
\underline{HD 239729}\\
 x $<$ 9.5 \micron$^{-1}$ &     0.484 &     0.642 &   4.342 & 0.937 &     4.597 &     1.219\\
 error   &     0.102 &     0.066 & 0.487 &     0.098 &     0.019 &     0.020\\
 x $<$ 8.7 \micron$^{-1}$ &     0.462 &     0.648 & 4.293 & 0.916 &     4.595 &     1.213\\
 error   &     0.106 &     0.067 &   0.484 & 0.103 &     0.016 &     0.020

\enddata 

\tablenotetext{a}{Parameters as defined in \citet{FM90}.}
\end{deluxetable} 

\newpage

\begin{deluxetable}{lrcrr} 
\tablecolumns{5}
\tablewidth{0pc} 
\tablecaption{Percentage of Available Carbon and Silicon used in 
Models\tablenotemark{a}}
\label{table_memfracs}
\tablehead{
\colhead{Sight line} & \colhead{Silicon} & \colhead{} &
\multicolumn{2}{c}{Carbon} \\
\cline{4-5}
\colhead{} & \colhead{AS\tablenotemark{b}} & \colhead{} & 
\colhead{AMC\tablenotemark{c}} & \colhead{Graphite}
}
\startdata 

\underline{HD 14250}\\
 x $<$ 9.5 \micron$^{-1}$  &   114\% &  &  51\% & 24\% \\
 
 x $<$ 8.7 \micron$^{-1}$ & 112\% &  & 51\% &  24\%\\
 
\underline{HD 62542}\\
 x $<$ 9.5 \micron$^{-1}$ &    99\% &  &   36\% &  8\% \\
 
 x $<$ 8.7 \micron$^{-1}$ &  90\% &  &   37\% &   7\% \\
 
\underline{HD 73882}\\
 x $<$ 9.5 \micron$^{-1}$ &     90\% &  &     53\% &     14\%\\
 
 x $<$ 8.7 \micron$^{-1}$ &   84\% & &    55\% &      12\%\\

\underline{HD 99872}\\
 x $<$ 9.5 \micron$^{-1}$ &  121\% &  &   76\% &    24\%\\

 x $<$ 8.7 \micron$^{-1}$ &     120\% & &     77\% &      24\% \\
 
\underline{HD 167971}\\
 x $<$ 9.5 \micron$^{-1}$  &   73\% &  &     54\% &    16\% \\
 
 x $<$ 8.7 \micron$^{-1}$  &     73\% & &     54\% &   16\% \\
 
\underline{HD 197770}\\
 x $<$ 9.5 \micron$^{-1}$ &     85\% &  &     32\% &  18\% \\
 
 x $<$ 8.7 \micron$^{-1}$ &     85\% &  &   35\% &  16\% \\

\underline{HD 210121}\\
 x $<$ 9.5 \micron$^{-1}$ &    120\% &  & 36\% &   14\% \\
 
 x $<$ 8.7 \micron$^{-1}$ &    119\% & &     36\% &     12\% \\

\underline{HD 239683}\\
 x $<$ 9.5 \micron$^{-1}$ &     78\% &  &     41\% &     16\% \\
 
 x $<$ 8.7 \micron$^{-1}$ &     77\% &  &     42\% &    15\% \\
 
\underline{HD 239729}\\
 x $<$ 9.5 \micron$^{-1}$ &     60\% &  &     27\% &     12\% \\
 
 x $<$ 8.7 \micron$^{-1}$ &   57\% &  &     28\% &     10\% \\

\enddata 

\tablenotetext{a}{The models were constrained to use a maximum of 120\% and
70\% of the available silicon and carbon along the sight lines 
as implied by the gas-to-dust ratios (see \S4.1). The only exception is 
HD 99872, which has an extinction curve that could not be fit within those
limits.}
\tablenotetext{b}{Astronomical silicates.}
\tablenotetext{c}{Amorphous carbon.}

\end{deluxetable}

\end{document}